# Semantic Retrieval at Walmart

Alessandro Magnani
Walmart Global Technology
Sunnyvale, USA
alessandro.magnani@walmart.com

Feng Liu
Walmart Global Technology
Sunnyvale, USA
f.liu@walmart.com

Suthee Chaidaroon*
Santa Clara University
Santa Clara, USA
schaidaroon@scu.edu

Sachin Yadav
Walmart Global Technology
Bangalore, India
Sachin.Yadav1@walmart.com

Praveen Reddy Suram
Walmart Global Technology
Bangalore, India
praveen.suram@walmart.com

Ajit Puthenputhussery
Walmart Global Technology
Sunnyvale, USA
ajit.puthenputhussery@walmart.com

Sijie Chen
Walmart Global Technology
Sunnyvale, USA
Sijie.Chen0@walmart.com

Min Xie*
Instacart
San Francisco, USA
min.xie@instacart.com

Anirudh Kashi*
University of Southern California
Los Angeles, USA
kashia@usc.edu

Tony Lee
Walmart Global Technology
Sunnyvale, USA
tony.lee@walmart.com

Ciya Liao
Walmart Global Technology
Sunnyvale, USA
ciya.liao@walmart.com

## ABSTRACT
In product search, the retrieval of candidate products before re-ranking is more critical and challenging than other search like web search, especially for tail queries, which have a complex and specific search intent. In this paper, we present a hybrid system for e-commerce search deployed at Walmart that combines traditional inverted index and embedding-based neural retrieval to better answer user tail queries. Our system significantly improved the relevance of the search engine, measured by both offline and online evaluations. The improvements were achieved through a combination of different approaches. We present a new technique to train the neural model at scale. and describe how the system was deployed in production with little impact on response time. We highlight multiple learnings and practical tricks that were used in the deployment of this system.

## 1 INTRODUCTION

Search is one of the most important channels for customers to discover products on an e-commerce website such as Walmart.com. Given our huge catalog which contains millions of products, helping users find relevant products for their queries is a very challenging problem [32]. Existing literature on information retrieval focuses mostly on web search [25]. While product search shares many common challenges with web search, there are many unique aspects of product search. Like other search, product search usually involves two steps: the first step is to *retrieve* all relevant products from the catalog that form the recall set; these candidate products then go through a *re-rank* step to identify which products are the best to return to the customer.

One major difference with web search is that the retrieval step in product search is a more critical and challenging problem [3, 5, 18]. This is because product titles (the main search-able text) are generally much shorter than web documents. Also, while many web documents may contain the same information, a specific product from a seller rarely has a duplicate. Retrieving a specific product while matching on shorter text is a challenging problem.

Traditionally, retrieval is based on text match between queries and documents, utilizing a heuristic score function like Okapi BM25 [25] and an inverted index [1] like Apache Lucene[1]. Text match between queries and documents suffers from vocabulary mismatch [23], which can be more problematic in product search [14, 32, 33]. For example, synonyms and hypernyms are difficult to handle [34]. Many existing works aim to solve this problem by incorporating knowledge graph [40] or having a dedicated query understanding component [6]. However, these approaches need a huge amount of domain expertise, and the cost of maintaining these components is high, since the catalog and product vocabulary frequently change in e-commerce.

---
[1]http://lucene.apache.org

More recently, neural retrieval systems have been proposed [14, 33] and deployed in production systems [43, 44] and have shown great success in bridging the vocabulary gap. However, neural systems are limited by the fact that the embedding size cannot be too large due to latency concerns. This is problematic when dealing with rare tokens [23].

In this paper, we describe the hybrid system used in production at Walmart.com and how it overcomes the individual limitations of traditional text-match retrieval and neural retrieval. We demonstrate the benefit of such a system for tail queries and highlight the learnings we had in the process of bringing the system to production. These include various challenges related to training the model, as well as engineering challenges in deploying the model to production while keeping the cost-to-serve low. We describe a solution that strikes a good balance between retrieval performance and model complexity in the context of product search.

The novelty of the paper are as follows:

- We deploy a hybrid search system on an e-commerce site with both inverted index and neural retrieval to handle high traffic of tail queries.
- We propose a novel method of selecting negative examples for training a large neural retrieval model and an approximate metric to evaluate the performance.
- We extensively explore the retrieval performance of a neural system across multiple dimensions, including model architecture, data preparation, and practical considerations.
- We report lessons learned and practical findings from developing and deploying the hybrid search system in the e-commerce website that serves millions of online customers daily.

## 2 RELATED WORK

*Neural information retrieval (NIR)* has been a popular topic in the search community recently. It leverages a set of sub-topics such as unsupervised learning of text embeddings like word2vec [26], deep Siamese [4] models based on query logs like CLSM [33] and DSSM [14], and query document interaction-based models like kernel pooling [39]. A good summary of the field can be obtained in [27]. In [23] the authors consider the tradeoff between sparse and dense retrieval and propose a multi embedding approach.

Multiple companies have described their production systems leveraging semantic retrieval. In [42], a two-tower neural model is trained using a mixed negative sampling in addition to batch random negatives. Baidu [44] described a production system that leverages multiple model pre-training strategies. In [13, 22] Facebook presented a system that combines an inverted index with semantic retrieval; the presented architecture includes multiple product features like images and titles. Taobao [43] proposed a way to better learn relevance from user engagement data. Sears used embeddings to represent products for a recommendation system [2]. In [29], Amazon presented a retrieval system based on a bag-of-words model; similar to our system, semantic retrieval was used in combination with a standard inverted index system. In [10] a residual-based learning framework was used to learn embeddings that compensate for shortcomings of the inverted index. In our approach the two systems are created separately.

*Negative item selection.* Multiple papers have investigated the problem of selecting negative samples to be used during training. Several works considered caching embeddings for the entire dataset [11, 41]. In [41], an iterative approach was suggested to find negative samples. We follow a similar approach, and extend it to improve the results in the common situation where not all relative items are known for a query. In [12], a streaming negative cache was used, but it cannot work for dual encoder training.

*Multiple embeddings.* There are also approaches where queries and items are represented by multiple embeddings for retrieval and ranking [8, 15, 20, 23, 24]. SPLADE [7] produces a sparse representation at the token level that improves storage requirements, and COIL [9] produces a token-level representation only for matching between query and document.

*Training strategies.* In [19], they showed for a question-answering task how a simple training strategy can effectively beat a state-of-the-art system. In [31], they used a teacher cross-interaction model, to help the training and selection of true negatives.

## 3 ARCHITECTURE

We propose a hybrid architecture that leverages the advantages of both traditional inverted index and neural retrieval. A traditional inverted index is still state of the art for retrieving documents with rare tokens [23] such as product ids and model numbers. Moreover, our production inverted index has capabilities like facet navigation and category filtering, which are hard to replicate using semantic retrieval alone. On the other hand, semantic retrieval helps bridge the vocabulary gap especially for tail queries; it helps with synonyms, misspellings, and other query variants that users type. Semantic retrieval also helps to better understand the semantics of longer queries that might contain a nuanced intent from the user.

The overall architecture is shown in Figure 1. When a user types a query, it is directed to the Query Planner to generate a query plan for inverted-index retrieval as well as a query embedding vector sent to the ANN Fetcher. The query embedding is then sent to an approximate nearest neighbor (ANN) index to retrieve the items with the closest embeddings. The ANN index contains the embeddings of all the products currently available in our catalog. When new products become available, a dedicated pipeline feeds the product information to the product embedding model to generate the embeddings. The embeddings are then stored in the updated ANN index.

Both the inverted index and ANN index retrieve a set of products which are merged to become the recall set. Finally, the retrieved items are ranked by our re-ranking system to produce the final list of products to be shown to the customers.

It is important to note that the semantic retrieval and inverted index are independently optimized. In this paper, we focus our attention on the semantic retrieval part and on some of the challenges of merging the two systems.

We discuss the semantic model in Section 4, the re-ranking model and its features in Section 6, and implementation details of the overall system in Section 7.

## 4 SEMANTIC MODEL

The semantic model architecture is a two tower structure as shown in Figure 2. Each tower produces an embedding for query and product respectively. The score of a query and product pair is the cosine similarity of the embeddings. We experimented with the inner product of the embedding as well (see Section 9).

The product information consists of a title, description, and a number of attribute values. Attribute values are, for example, color, brand, material. Attributes are not always available for all products. We experimented with different number of attributes in Section 8.

There are two main classes of model used. The first one is a traditional bag of words model [16, 29] (BoW) and the second is based on a BERT [35] architecture. The BERT based model is far superior for this application and it will be our main focus. We report the performance of the BoW model as a baseline.

### 4.1 Models

We have experimented with different transformer architecture by leveraging HuggingFace pre-trained models repository [2]. Specifically we have used the *BaseBERT* with 12 layers and 1024 embedding size and *DistillBERT* with 6 layers and 768 embedding size. We report below on the performance of different model architectures. We use the pre-trained tokenizer, and all training is done by starting with the pre-trained model. Our experiments use identical towers (Siamese network). For most experiments, we use the embedding vector corresponding to the special token "[CLS]".

In our experiments in Section 8, titles provide most of the signal for retrieval. We use a baseline model with only title as input. We then added more attributes to the input. Each attribute is concatenated to the title by using a prefix which is an unused token selected specifically for the attribute. For example, when adding the color red to a product, the input looks like "[title tokens] [color token] red". This technique allows the model to determine which attributes have been concatenated. We experiment with four common attributes (product category, brand, color and gender) and with a longer list of 26 attributes including the basic ones.

### 4.2 Loss function

We used a sampled softmax loss where for each query we have both relevant and irrelevant products with a corresponding score. As described in section 5.2, selecting negative items is essential for good performance. We also notice that allowing multiple relevant products during training helps. Since there are, in general, many relevant products for a given query, we sample a few relevant products for each epoch.

$$loss_i = \sum_j^N S_{ij} \log \frac{\exp\left(\cos\left(q_i, p_j\right)/\sigma\right)}{\sum_j^N \exp\left(\cos\left(q_i, p_j\right)/\sigma\right)} \quad (1)$$

Equation 1 shows the loss contribution of query $i$, where $N$ is the number of products under consideration, $S_{ij}$ is the score of product $j$ for query $i$, and $q_i$, $p_j$ are the embedding for query $i$ and product $j$ respectively. $\sigma$ is a temperature factor that is trained together with all the model parameters. Other loss functions have been evaluated including pointwise and pairwise losses, but the softmax loss outperforms them and is very robust during training. The number of products considered for each query is $N = 20$. There is a trade-off between $N$ and the batch size. Increasing $N$ reduces the batch size and therefore also the in-batch negatives.

We will discuss in Section 5 how the scores $S_{ij}$ are selected and how the products are selected in more detail.

## 5 DATA

The training of the model is performed on engagement data collected at Walmart.com over a one year period. The data contains the top 2 million queries based on number of impressions. For each query, we consider the products that were shown to the customers, and construct labels based on the corresponding number of purchases, clicks, and impressions. Note that we do not account for presentation bias, since in a retrieval model, the order in which items are returned is not relevant. In our experience correcting for presentation bias adds more complexity for a negligible performance improvement.

### 5.1 Labeling

For each query and product pair we assign a score $S_{ij}$ between 0 and 10. Since the loss is insensitive up to a multiplicative factor, the range of score can be selected arbitrarily. Query-product pairs with purchases are assigned the highest scores between 10 and 8. Products that only received clicks are assigned scores between 7 and 5, and if products only received impressions scores between 4 and 2. We assigned scores of 0 to *negative* items as described in section 5.2.

Ordered products are assigned a score based on a smoothed estimation of their order rate $rate = \frac{orders+\alpha}{impressions+\alpha}$ where $\alpha$ is the smoothing factor. The product with the highest order rate receives a score of 10, while the product with the lowest order rate receives a score of 8 following equation $S = (10 - 8)\frac{rate - \min(rate)}{\max rate - \min rate} + 8$. A similar approach is used for products that received only clicks.

Although this labeling strategy is arbitrary, it has been shown to be effective in practice. In particular differentiating the scores of items that have been purchased from items that have only been clicked is in our experience essential to create effective retrieval models. This is consistent with [29] even though we use a different loss.

### 5.2 Negative item selection

The selection of negative items is necessary to help the model discern a relevant title among millions of products. We used two sources of negative products: random products from within a batch

---
[2]https://huggingface.co/models

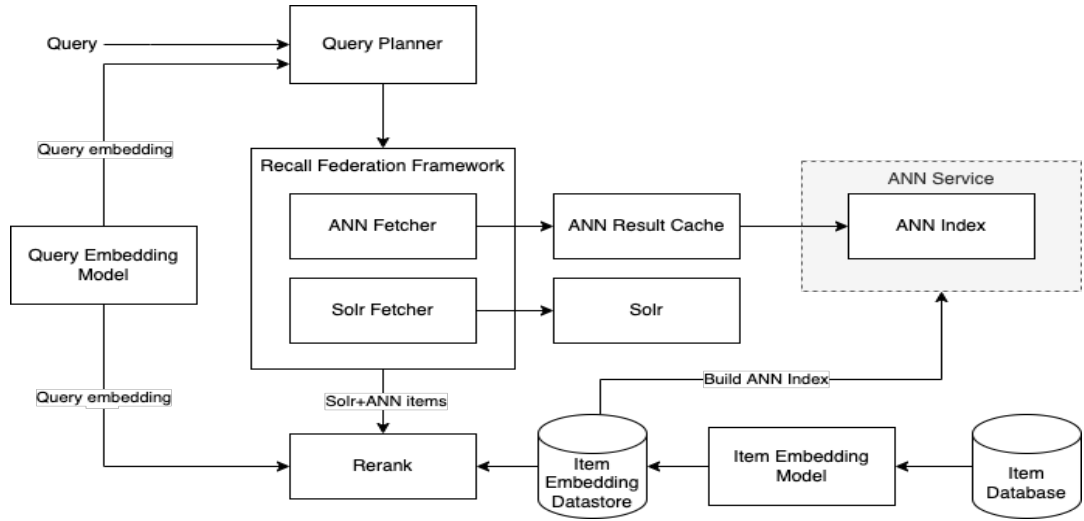

Figure 1: The system architecture of hybrid retrieval system

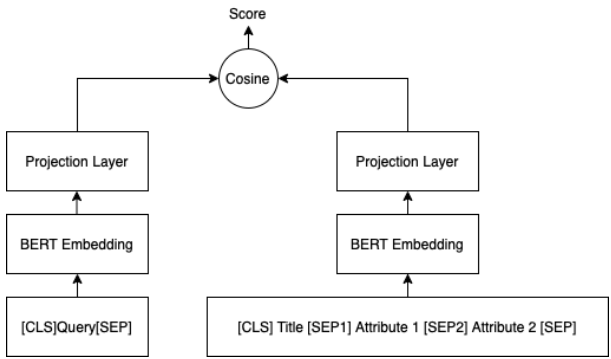

Figure 2: The model architecture

during training and *hard negatives* selected offline for a given trained model.

*5.2.1 In batch negatives.* Selecting negative items from a batch is a common technique that reduces the computation because the embeddings are already computed for all the products in the batch. Since for memory constraints, it is not possible to select all products in a batch as negative, we experimented with two ways of sampling them. The first approach is a *random selection*. The second approach is to perform a *hard in batch selection* [22]. This approach focuses on only the hardest samples in a training batch. For a given query in a batch the negatives are generated by using all the products from other queries in the batch as a pool and selecting only the ones which receive the highest cosine similarity.

*5.2.2 Hard negative search.* Out of the several negative construction approaches available, we used ANCE [41] to which we added novel strategies that cause the model to retrieve more relevant results, thereby boosting the recall metric. Unlike in the original paper [41] where the relevant items are known, in the e-commerce setting, there can be many relevant products for a given query. For example, a query like *"red shoes"* can have hundreds of relevant products. Since it is impractical to editorially evaluate all of them, we introduce a set of different heuristics to overcome this limitation.

The procedure to generate hard negatives is as follows: 1) Generate top-k results for each query in training corpus using a partially trained model 2) Select negatives from top-k results based on a selection strategy (see below) 3) inject the hard negatives into the training corpus and resume training 4) repeat the steps 1-3.

We explored three strategies to select negatives out of the top-k results.

- *PT match*: For each item in the top-k results of a given query, if the item is not in the training data and matches the product type (PT) attribute of the top-m items in the training data, it is removed from the set of negative candidates.
- *Token match*: The negative candidates generated using this strategy are more strict. We take PT match candidates as an input to this and for each candidate item find its overlap with query tokens. If the overlap score is below threshold $t$, we keep the item; otherwise discard it. We have experimented with different thresholds and found that $t$=0.5 gave best results.
- *Student-Teacher*: We trained a separate model (Teacher) wherein the query tokens can directly attend to product tokens. The teacher model is a MonoBERT-based [38] single encoder network that concatenates query and the product information together as input. The top-k items are generated via this model, and the above PT match filter is applied to get the final negative candidates that are then injected into the embedding model (Student).

We will compare these three approaches in the result section.

## 5.3 Reducing the model size

The initial embedding size was 768 for which the best recall was obtained in our offline evaluations. This embedding size creates a fairly large storage footprint for the ANN index (Section 7.1) and item embedding (Section 6). Due to system-level limitation, item embedding storage is updated through a publish-subscribe based messaging pipeline that does not accept partial updates. Reducing the size of embedding is beneficial as it allows the item embedding and the ANN index to be refreshed more frequently. The cost for updating the ANN index, which depends on the storage size of the input, is also taken into account.

We therefore investigated two different strategies to reduce the size of the embeddings. In the first approach, we added a linear projection layer to reduce the embedding size to 368, 256, 128, and 64. In the second approach we used a transformer architecture that has a smaller embedding size. We specifically picked the MiniLM [36] architecture with 12 layers and an embedding size of 368 and the XtremeDistil [28] architecture with 6 layers and an embedding size of 368. In both cases, the implementation and the checkpoint were provided by Huggingface [37]. As shown in the result Section 8, the linear projection is very effective in reducing the size of the embedding with very little performance cost.

### 5.4 Multi Embeddings

E-commerce queries often have multiple interpretations. For example, the query *"apple"* could refer to the fruit or to the electronics brand. For this reason, we explore the possibility of having multiple embeddings to represent a query or product. For a query, this means to generate $m$ embeddings, and make $m$ calls to the ANN service. Denoting $q_i^d$ as the $d$th embedding of the $i$th query, the product score is $\max\left(\cos(q_i^d, p_j)\right)$ for $d \in 1 \ldots m$. Following [23], we let the $m$ embeddings correspond to the first $m$ tokens. This is a natural extension of using the embedding corresponding to the first token "[CLS]". For multiple embeddings on the product side, like in [23], there are $n$ embeddings, and we therefore store in the ANN service $n$ times the number of products.

### 5.5 Freezing token embeddings

We also experimented with a different setup where the token embeddings are kept frozen to preserve the learning of the pretrained model. The hypothesis is that the pretrained model which has been trained on a much larger dataset, can better preserve its learning if the embeddings are not changed.

### 5.6 Different pooling

The last experiments are regarding the pooling mechanism on top of the BERT model. We tried three different options. The first one uses the output of the [CLS] token. The second uses the average of the embeddings coming from all tokens. The third uses the maximum per component across all tokens.

## 6 RE-RANK STAGE

In this section, we describe how the two recall sets from the inverted index and semantic retrieval are merged and ranked. A key aspect of e-commerce search is the presence of useful ranking features that capture the different query and product attributes. The features used in the ranking model can be organized into the following groups:

- Query Features: These features capture the different attributes and properties learnt from the query. For example, query attributes like product type, brand, query length.
- Item Features: These features capture the different product attributes as well as engagement features computed at the product level. For example, title attributes, title length, user ratings, user reviews, product sales, product department, etc.
- Query-Item Features: These features capture the relations related to the query-item pair. For example, BM25 text match score, query-item engagement, query-item attribute match score.

Figure 1 also shows the re-rank architecture with the query-product BERT embedding feature. The query is encoded to the query embedding at runtime. The item embedding for all the items in the catalog are pre-generated and saved to the item embedding datastore. For all the items from both the inverted index and ANN, the corresponding item embeddings are fetched from the item embedding datastore.

At the re-rank layer, the cosine similarity between query and item embeddings is included as a query-item feature. The re-rank model, which is a Gradient Boosted Decision Tree (GBDT) model, ranks all items in the merged list.

## 7 IMPLEMENTATION

In this section we discuss some of the engineering considerations and challenges in implementing the hybrid retrieval system in production.

### 7.1 ANN service

One major challenge for online neural product retrieval is the trade-off among accuracy, speed, and memory. Brute force algorithms that retrieve the exact $k$ closest vectors of a query vector from millions of item vectors with respect to a pre-defined distance cannot be used in the production setup due to their high time complexity. Some amelioration can be obtained by first compressing the data size so that it may be easier for the vectors to be fed into memory. Such techniques include but are not limited to locality sensitive hashing (LSH), quantization, and product quantization (PQ). For faster retrieval, not all indices will be scanned when a query is executed. Candidate vectors are usually split into multiple clusters, and only vectors in the closest few clusters will be scanned - if not further reduced by other pruning methods.

Tools and services that support approximate nearest neighbor (ANN) search have emerged in the past few years. One popular tool among developer communities and has the potential to be productionized is FAISS[3]. However, to reduce the cost of maintenance and to minimize the system level risk, as the real traffic might surge during certain periods of time, we use a managed ANN service available commercially.

As with all other ANN algorithms, hyperparameter tuning is necessary to achieve the desired recall quality within an acceptable level of latency, storage, and computation cost. Our experiments show that with normalized vectors of dimension 256, the ANN services can yield 99% for recall@20, evaluated against the full nearest neighborhood search, with an average latency around 13 ms;

---
[3]https://ai.facebook.com/tools/faiss/

| d256 embedding | | | d768 embedding | | |
|---|---|---|---|---|---|
| Recall@20 | Latency | Storage | Recall@20 | Latency | Storage |
| 97.44% | 6.5 ms | | 95.65% | 6.7 ms | |
| 98.22% | 8.1 ms | 100% | 97.74% | 8.9 ms | 300% |
| 98.94% | 12.8 ms | | 98.84% | 12.9 ms | |

Table 1: The recall@20, latency and storage of the embedding of 256 and 768 dimensions.

| Model serving variation | P99 latency increased |
|---|---|
| BERT 6-layer | + 100.0% |
| BERT 2-layer | + 97.88% |
| BERT 2-layer (custom lookup implementation) | + 50.26% |
| **BERT 2-layer (custom lookup implementation, fixed input shape)** | **+ 30.14%** |

Table 2: Impact on latency of the query understanding module

for normalized vectors of dimension 768, the services can achieve a similar recall@20 but with three times the storage space. This implies a much higher cost of running the system and justify the decision to decrease the embedding size.

Table 1 reports the recall-latency trade-off for a few combinations of hyperparameters on the ANN service.

## 7.2 Runtime Implementation

The in-house search engine accepts queries from customers. If a query is eligible for neural retrieval (i.e. a tail query), the query planner sends its embedding vector, which is the output of the query encoder, to the ANN index. The results of the ANN index are cached with a preset time-to-alive (TTL) to reduce latency and cost.

The recall federation framework retrieves products from both the ANN index and the inverted index, and then de-duplicates and merges the product sets before sending them to the rerank phase.

## 7.3 Query modeling latency and consideration

In real production scenarios, a large portion of queries submitted by users are not predictable and hence cannot be vectorized offline beforehand to reduce the overall runtime latency. For simple models such as the Bag-of-Words (BoW) model, the computation is fast and usually does not raise any concern. However, the power of such simple models is also limited for this application. In contrast, BERT-based models describe in section 4 and their variations have drawn a lot of attentions to their capacity in semantic understanding, as well as to its application in solving search ranking problems, and have shown in our experiments there much higher performance along multiple metrics. Unfortunately, BERT-based models also require more computation resources and may increase response time of the runtime system.

We integrate the BERT-based query encoder as part of our query understanding module. We found that with the same capacity of computing clusters with CPUs, the 6-layer Distilled BERT model almost doubled the P99 latency of the query understanding module. Our original plan was to reduce the number of layers from 6 to 2. However, we found that the latency was not reduced linearly with the number of layers and the contribution of the token embedding lookup was substantial.

Our model is exported from Torch checkpoints into ONNX[4] format and is served in a Java codebase. The embedding lookup operation implementation in the ONNX backend seems to be highly inefficient. Therefore, in our experiments, we tried to move this embedding lookup to Java *hashmap* and to feed the model with gathered token embedding matrices. Furthermore, we found that setting a dynamic query length was causing extra latency. For this reason, we opted for fixing the maximum length of the query and using padding. As shown in Table 2, this improve latency even further.

[4]https://github.com/onnx/onnx

The impact of different model variations on the latency of our query understanding module are logged in Table 2, with the original BERT 6-layer model normalized as 100%.

To further minimize the latency impact, we deployed the models to individual computing clusters with GPUs for remote serving. Using servers with 4 cores of Nvidia Tesla T4 GPU, we managed to eliminate the extra latency introduced by our 6-layer Distilled BERT model on top of our existing query understanding module.

## 8 EXPERIMENTS AND RESULTS

All modeling effort has been performed using a dataset of 2 million queries collected over a period of one year from Walmart.com logs. The dataset was divided between a training and validation dataset of size 90% and 10% respectively. The test dataset contains 140 thousand queries disjoint from both training and validation for which a set of relevant items has been identified using user engagement and editorial feedback.

The training was performed using PyTorch [30] and Hugging-Face. Adam [21] was used to train with a learning rate of $10^{-5}$. The batch size was 40, and the number of products during training for each query was set to 20.

### 8.1 Offline Metrics

We evaluate the models using a *recall metric* which measures the percentage of relevant products retrieved for a golden dataset. The golden dataset contains a set of relevant products for 140 thousand queries out of a set of around 7 million products. Since scoring all products for a given query is not possible, only a subset of the relevant products is known.

It has been clear that the recall metric alone does not fully capture the performance of the semantic search. We noticed that the model could have a relatively high recall while at the same time retrieving a set of completely unrelated products. This is in part due to the small user engagement available for some products as well as the presence in the catalog of products with noisy titles. For this reason, we define a new metric that tries to measure the approximate number of irrelevant products. Since in e-commerce, each product has a product category associated with it (e.g. dining chairs, toothpaste), we make the simplifying assumptions that all relevant products for a query should have the same product category of at least one of the relevant products in the golden dataset. This is the same assumption made in our approach to select negative items in Section 5.2. The metric that we call *category recall*, is defined as the percentage of products in the recall set that has the same product category of at least one product in the golden dataset. Clearly, this is not a perfect metric because often there could be more product categories that are relevant for a query and not all of them are represented in the golden dataset. Moreover, product category can often be incorrectly assigned adding noise to this metric. Nevertheless, we have found this metric very useful in driving our modeling effort and capable of capturing what our manual inspection had found. During training, we compute the recall metric only on the batch as a proxy for the overall recall, and we use it to terminate the training.

| Model | Recall@40 | Cat Recall@40 |
|---|---|---|
| Inverted Index | 0% | **0%** |
| BoW 256 | -7.3% | - |
| Siamese BERT 12-layer 768 | +8.25% | - |
| Siamese BERT 6-layer 1024 | +10.12% | - |
| Siamese DBERT 6-layer 768 | +12.33% | -18.33% |
| Siamese DBERT 6-layer 768 + (product cat., brand, color, gender) | **+18.22%** | -16.83% |
| Siamese DBERT 6-layer 768 + description + 25 other attributes | +16.34% | -16.78% |

**Table 3: Offline model performance by number of layers and embedding size, with random negatives**

| Model | Recall@40 | Cat Recall@40 |
|---|---|---|
| Siamese DBERT 6-layer 768 (+random negatives) | 0% | 0% |
| Siamese DBERT 6-layer 768 (+PT match filter negatives) | +0.87% | +18.21% |
| Siamese DBERT 6-layer 768 (+PT match + token match filter negatives) | +2.86% | +14.08% |
| Siamese DBERT 6-layer 768 (+teacher student negatives) | -0.12% | +14.93% |
| Siamese DBERT 6-layer 768 (+PT match + token match filter negatives + in-batch hard negatives) | **+5.15%** | **+20.47%** |

**Table 4: Offline model performance for different negative selection techniques**

| Model | Recall40 | Cat Recall@40 |
|---|---|---|
| Siamese DBERT 6-layer 768 (baseline) | 0% | **0%** |
| Siamese DBERT 6-layer 768 (linear 368) | **+0.75%** | -0.68% |
| Siamese DBERT 6-layer 768 (linear 256) | +0.18% | -0.23% |
| Siamese DBERT 6-layer 768 (linear 128) | -4.23% | -0.64% |
| Siamese DBERT 6-layer 768 (linear 64) | -17.48% | -2.94% |
| MiniLM 12-layer 368 | -12.39% | -6.25% |
| XtremeDistil 6-layer 368 | -18.28% | -7.72.% |

**Table 5: Offline model performance for different embedding size reduction**

## 8.2 Offline model results

In this section, we report the results of our modeling effort based on the two metrics described in Section 8.1. All the results are reported with respect to a baseline model specified for each set of experiments. We also report as baseline the performance of a simple inverted index implementation where only the title of the product is indexed. In Table 3, we report all the main findings with respect to number of layers and embedding size. We observed that the BoW model had lower offline numbers compared to inverted index lookup. After switching to BERT based model, we were able to beat the baseline by 8.25% in Recall@40 and switching to DistilBERT model gave an extra lift of 4%. We experimented with including product attributes, and got a further 6% boost after including attributes like product category, brand, color and gender to the input for a product. But, when we added even more product attributes (like description etc.), we did not see any further improvement in model performance. Notice how Category Recall@40 is dramatically lower without the use of negatives.

In Table 4, the performance of different negative selection techniques are shown. The first row corresponds to the best model in Table 3. When we added hard negatives to the training data, described in section 5.2, we observed 0.87% lift in Recall@40 and 18% lift Category Recall@40 when using only product category match. When combining that with token match, the improvement on recall is 2.8%. On the other hand, our student-teacher negative selection does not work as well as the other approaches. Finally, adding hard in-batch negatives improved the recall by around 5%.

In Table 5 we show the results of the different embedding size reductions described in section 5.3 The linear projection matrix performs at a very similar performance than the original model up to a size of 256 before dropping in performance. Moreover, at a similar embedding size, a projection layer is superior to architecture that have smaller sizes like the MiniLM. Therefore, 256 is the size used in production.

Table 6 shows the results of the multiple embeddings using 3 embeddings. As we can see the multiple embedding doesn't seem to bring any advantage over a single embedding in our experiments. Considering the higher cost of the solution, it has not been implemented.

Table 7 shows that freezing the token embeddings during training provides a small improvement and seems to confirm the idea that the model can better generalize. Finally Table 8 shows the effect of different pooling approaches. The average pooling and default pooling on [CLS] tokens have almost identical performance. Max pooling has the worst performance.

| Model | Recall@40 | Cat Recall@40 |
|---|---|---|
| DBERT 768 | **0%** | **0%** |
| Multi embedding - query (3 * 768) | -3.9% | -1.13% |
| Multi embedding - product (3 * 768) | -2.69% | -2.11% |

Table 6: Offline model performance for multiple query and product embeddings

| Model | Recall@40 | Cat Recall@40 |
|---|---|---|
| Siamese BERT 6-layer 768 | 0% | 0% |
| Siamese BERT frozen tokens | +1.75% | +1.21% |

Table 7: Performance of model with frozen token embeddings.

| Model | Recall@40 | Cat Recall@40 |
|---|---|---|
| [CLS] pooling | **0%** | **0%** |
| Average Pooling | -0.28% | -0.01% |
| Max Pooling | -9.34% | -0.04% |

Table 8: Performance of the Siamese DBERT 6-layer 768 model with different pooling.

| Method | NDCG@5 Lift (P-value) | NDCG@10 Lift (P-value) |
|---|---|---|
| BERT Embedding with dimension 768 (offline hard negatives) | +1.42% (0.09) | +2.88% (0.00) |
| BERT Embedding with dimension 256 (offline hard negatives + in-batch hard negatives) | +2.02% (0.03) | +2.84% (0.00) |

Table 9: Human evaluation on the top-10 ranking results on a random sample of tail search traffic queries by the proposed architecture.

| Method | ATC@40 Lift (P-value) |
|---|---|
| BERT Embedding with dimension 768 (offline hard negatives) | +0.41% (0.00) |
| BERT Embedding with dimension 256 (offline hard negatives + in-batch hard negatives) | +0.54% (0.00) |

Table 10: Interleaving results on the top-40 ranking for the proposed architecture.

## 8.3 Live Experiments

*8.3.1 Manual Evaluation Results.* We evaluated the performance of our proposed architecture by using human assessors to evaluate the top-10 ranking results of the proposed architecture and the current production at Walmart. The candidate model uses an embedding size of 256 and uses the "[CLS]" pooling. For a query, the human assessors are shown the product image, title and price of the product along with the product link in Walmart website. They rate the relevance of the product on a 3-point grading scale as not relevant, relevant with missing attributes, and perfect match. The queries are selected via random sample of the search traffic from the tail segment. As shown in Table 9, the proposed architecture significantly improves the relevance of tail queries. Note that BERT embedding with dimension 256 performed similarly or even better than dimension 768.

*8.3.2 Interleaving Results.* We assessed the user engagement performance of our proposed architecture compared with the current production at Walmart using interleaving [17]. We evaluate two different models that have the same DistillBERT architecture but in one case they have a linear projection layer to an embedding size of 256. Interleaving is an online evaluation approach where each user is presented with a combination of ranking results from both the control and variation. We observe the add-to-carts (ATC) between the control and variation ranking. The metric measured is ATC@40 which is the count of add-to-carts in the top 40 position between control and variation ranking models. The results shown in Table 10 demonstrate the effectiveness of the proposed architecture in improving the user engagement performance. We notice a similar pattern in interleaving results, where the BERT embedding with dimension 256 performs better than dimension 768.

## 9 LESSONS LEARNED

Among the many things that we learned while creating this system, we would like to highlight a few of them.

*Inner product vs. cosine similarity.* During the model training, we experimented with inner product. The inner product is more stable during training and does not require the temperature factor $\sigma$ shown in Eq. 1. This removes the need to select $\sigma$ and in general produces better results. Unfortunately, inner product was much harder to optimize when creating the ANN index, compared to cosine similarity. For this reason, we eventually focused on cosine similarity only.

*Blending features.* Many text fields are generally available for each product. These include different descriptions and many attributes. There was a common belief that the description of the product would help improve the recall performance. In all our experiments, we could not extract any boost in performance. This is probably because descriptions can contain a lot of irrelevant text that simply adds noise.

*Model complexity.* As seen in the Section 8, for this application, a very deep model or very large embedding size is not necessary to achieve top performance. This is probably because queries and product titles are not very complex from a semantic perspective.

## REFERENCES
[1] Artem Babenko and Victor Lempitsky. 2014. The inverted multi-index. *IEEE transactions on pattern analysis and machine intelligence* 37, 6 (2014), 1247–1260.

[2] Bibek Behera, Manoj Joshi, Abhilash KK, and Mohammad Ansari Ismail. 2017. Distributed Vector Representation Of Shopping Items, The Customer And Shopping Cart To Build A Three Fold Recommendation System. *arXiv preprint arXiv:1705.06338* (2017).
[3] Eliot Brenner, Jun Zhao, Aliasgar Kutiyanawala, and Zheng Yan. 2018. End-to-End Neural Ranking for eCommerce Product Search: an application of task models and textual embeddings. In *eCom*.
[4] Jane Bromley, James W Bentz, Léon Bottou, Isabelle Guyon, Yann LeCun, Cliff Moore, Eduard Säckinger, and Roopak Shah. 1993. Signature verification using a "siamese" time delay neural network. *IJPRAI* 7, 04 (1993), 669–688.
[5] Huizhong Duan, ChengXiang Zhai, Jinxing Cheng, and Abhishek Gattani. 2013. Supporting Keyword Search in Product Database: A Probabilistic Approach. *PVLDB* 6, 14 (2013), 1786–1797.
[6] Susan T. Dumais. 2016. Personalized Search: Potential and Pitfalls. In *CIKM, 2016*. 689.
[7] Thibault Formal, Benjamin Piwowarski, and Stéphane Clinchant. 2021. SPLADE: Sparse lexical and expansion model for first stage ranking. In *SIGIR*. 2288–2292.
[8] Luyu Gao, Zhuyun Dai, and Jamie Callan. 2020. Modularized transformer-based ranking framework. *arXiv preprint arXiv:2004.13313* (2020).
[9] Luyu Gao, Zhuyun Dai, and Jamie Callan. 2021. COIL: Revisit Exact Lexical Match in Information Retrieval with Contextualized Inverted List. *arXiv preprint arXiv:2104.07186* (2021).
[10] Luyu Gao, Zhuyun Dai, Tongfei Chen, Zhen Fan, Benjamin Van Durme, and Jamie Callan. 2021. Complement lexical retrieval model with semantic residual embeddings. In *European Conference on Information Retrieval*. Springer, 146–160.
[11] Kelvin Guu, Kenton Lee, Zora Tung, Panupong Pasupat, and Ming-Wei Chang. 2020. Realm: Retrieval-augmented language model pre-training. *arXiv preprint arXiv:2002.08909* (2020).
[12] Kaiming He, Haoqi Fan, Yuxin Wu, Saining Xie, and Ross Girshick. 2020. Momentum contrast for unsupervised visual representation learning. In *CVPR*. 9729–9738.
[13] Jui-Ting Huang, Ashish Sharma, Shuying Sun, Li Xia, David Zhang, Philip Pronin, Janani Padmanabhan, Giuseppe Ottaviano, and Linjun Yang. 2020. Embedding-based retrieval in facebook search. In *Proceedings of the 26th ACM SIGKDD International Conference on Knowledge Discovery & Data Mining*. 2553–2561.
[14] Po-Sen Huang, Xiaodong He, Jianfeng Gao, Li Deng, Alex Acero, and Larry P. Heck. 2013. Learning deep structured semantic models for web search using clickthrough data. In *CIKM, 2013*. 2333–2338.
[15] Samuel Humeau, Kurt Shuster, Marie-Anne Lachaux, and Jason Weston. 2019. Poly-encoders: Transformer architectures and pre-training strategies for fast and accurate multi-sentence scoring. *arXiv preprint arXiv:1905.01969* (2019).
[16] Thorsten Joachims. 1998. Text categorization with support vector machines: Learning with many relevant features. In *ECML*. 137–142.
[17] Thorsten Joachims et al. 2003. Evaluating Retrieval Performance Using Clickthrough Data.
[18] Shubhra Kanti Karmaker Santu, Parikshit Sondhi, and ChengXiang Zhai. 2017. On application of learning to rank for e-commerce search. In *Proceedings of the 40th international ACM SIGIR conference on research and development in information retrieval*. 475–484.
[19] Vladimir Karpukhin, Barlas Oğuz, Sewon Min, Patrick Lewis, Ledell Wu, Sergey Edunov, Danqi Chen, and Wen-tau Yih. 2020. Dense passage retrieval for open-domain question answering. *arXiv preprint arXiv:2004.04906* (2020).
[20] Omar Khattab and Matei Zaharia. 2020. Colbert: Efficient and effective passage search via contextualized late interaction over bert. In *SIGIR*. 39–48.
[21] Diederik P Kingma and Jimmy Ba. 2014. Adam: A method for stochastic optimization. *arXiv preprint arXiv:1412.6980* (2014).
[22] Yiqun Liu, Kaushik Rangadurai, Yunzhong He, Siddarth Malreddy, Xunlong Gui, Xiaoyi Liu, and Fedor Borisyuk. 2021. Que2search: Fast and accurate query and document understanding for search at facebook. In *Proceedings of the 27th ACM SIGKDD Conference on Knowledge Discovery & Data Mining*. 3376–3384.
[23] Yi Luan, Jacob Eisenstein, Kristina Toutanova, and Michael Collins. 2021. Sparse, Dense, and Attentional Representations for Text Retrieval. *Transactions of the Association for Computational Linguistics* 9 (2021), 329–345.
[24] Sean MacAvaney, Franco Maria Nardini, Raffaele Perego, Nicola Tonellotto, Nazli Goharian, and Ophir Frieder. 2020. Efficient document re-ranking for transformers by precomputing term representations. In *SIGIR*. 49–58.
[25] Christopher D. Manning, Prabhakar Raghavan, and Hinrich Schütze. 2008. *Introduction to information retrieval*. Cambridge University Press.
[26] Tomas Mikolov, Kai Chen, Greg Corrado, and Jeffrey Dean. 2013. Efficient Estimation of Word Representations in Vector Space. *CoRR* abs/1301.3781 (2013).
[27] Bhaskar Mitra and Nick Craswell. 2018. An Introduction to Neural Information Retrieval. *Foundations and Trends in Information Retrieval* 13, 1 (2018), 1–126.
[28] Subhabrata Mukherjee and Ahmed Awadallah. 2020. XtremeDistil: Multi-stage distillation for massive multilingual models. *arXiv preprint arXiv:2004.05686* (2020).
[29] Priyanka Nigam, Yiwei Song, Vijai Mohan, Vihan Lakshman, Weitian Ding, Ankit Shingavi, Choon Hui Teo, Hao Gu, and Bing Yin. 2019. Semantic product search. In *Proceedings of the 25th ACM SIGKDD International Conference on Knowledge Discovery & Data Mining*. 2876–2885.
[30] Adam Paszke, Sam Gross, Francisco Massa, Adam Lerer, James Bradbury, Gregory Chanan, Trevor Killeen, Zeming Lin, Natalia Gimelshein, Luca Antiga, Alban Desmaison, Andreas Kopf, Edward Yang, Zachary DeVito, Martin Raison, Alykhan Tejani, Sasank Chilamkurthy, Benoit Steiner, Lu Fang, Junjie Bai, and Soumith Chintala. 2019. PyTorch: An Imperative Style, High-Performance Deep Learning Library. In *Advances in Neural Information Processing Systems 32*, H. Wallach, H. Larochelle, A. Beygelzimer, F. d'Alché-Buc, E. Fox, and R. Garnett (Eds.). Curran Associates, Inc., 8024–8035. http://papers.neurips.cc/paper/9015-pytorch-an-imperative-style-high-performance-deep-learning-library.pdf
[31] Yingqi Qu, Yuchen Ding, Jing Liu, Kai Liu, Ruiyang Ren, Wayne Xin Zhao, Daixiang Dong, Hua Wu, and Haifeng Wang. 2021. RocketQA: An optimized training approach to dense passage retrieval for open-domain question answering. In *NAACL-HLT*.
[32] Fatemeh Sarvi, Nikos Voskarides, Lois Mooiman, Sebastian Schelter, and Maarten de Rijke. 2020. A Comparison of Supervised Learning to Match Methods for Product Search. *SIGIR Workshop on eCommerce* (2020).
[33] Yelong Shen, Xiaodong He, Jianfeng Gao, Li Deng, and Grégoire Mesnil. 2014. A Latent Semantic Model with Convolutional-Pooling Structure for Information Retrieval. In *CIKM, 2014*. 101–110.
[34] Yelong Shen, Xiaodong He, Jianfeng Gao, Li Deng, and Grégoire Mesnil. 2014. Learning semantic representations using convolutional neural networks for web search. In *Proceedings of the 23rd international conference on world wide web*. 373–374.
[35] Ashish Vaswani, Noam Shazeer, Niki Parmar, Jakob Uszkoreit, Llion Jones, Aidan N. Gomez, Lukasz Kaiser, and Illia Polosukhin. 2017. Attention is All you Need. In *NeurIPS*. 6000–6010.
[36] Wenhui Wang, Furu Wei, Li Dong, Hangbo Bao, Nan Yang, and Ming Zhou. 2020. Minilm: Deep self-attention distillation for task-agnostic compression of pre-trained transformers. *NeurIPS* 33 (2020), 5776–5788.
[37] Thomas Wolf et al. 2020. Transformers: State-of-the-Art Natural Language Processing. In *EMNLP*.
[38] Ji Xin, Rodrigo Nogueira, Yaoliang Yu, and Jimmy Lin. 2020. Early exiting BERT for efficient document ranking. In *Proceedings of SustaiNLP: Workshop on Simple and Efficient Natural Language Processing*. 83–88.
[39] Chenyan Xiong, Zhuyun Dai, Jamie Callan, Zhiyuan Liu, and Russell Power. 2017. End-to-end neural ad-hoc ranking with kernel pooling. In *Proceedings of the 40th International ACM SIGIR conference on research and development in information retrieval*. 55–64.
[40] Chenyan Xiong, Russell Power, and Jamie Callan. 2017. Explicit Semantic Ranking for Academic Search via Knowledge Graph Embedding. In *WWW, 2017*. 1271–1279.
[41] Lee Xiong, Chenyan Xiong, Ye Li, Kwok-Fung Tang, Jialin Liu, Paul Bennett, Junaid Ahmed, and Arnold Overwijk. 2020. Approximate nearest neighbor negative contrastive learning for dense text retrieval. *arXiv preprint arXiv:2007.00808* (2020).
[42] Ji Yang, Xinyang Yi, Derek Zhiyuan Cheng, Lichan Hong, Yang Li, Simon Xiaoming Wang, Taibai Xu, and Ed H Chi. 2020. Mixed negative sampling for learning two-tower neural networks in recommendations. In *Companion Proceedings of the Web Conference 2020*. 441–447.
[43] Shaowei Yao, Jiwei Tan, Xi Chen, Keping Yang, Rong Xiao, Hongbo Deng, and Xiaojun Wan. 2021. Learning a Product Relevance Model from Click-Through Data in E-Commerce. In *Proceedings of the Web Conference*. 2890–2899.
[44] Lixin Zou, Shengqiang Zhang, Hengyi Cai, Dehong Ma, Suqi Cheng, Shuaiqiang Wang, Daiting Shi, Zhicong Cheng, and Dawei Yin. 2021. Pre-trained language model based ranking in Baidu search. In *Proceedings of the 27th ACM SIGKDD Conference on Knowledge Discovery & Data Mining*. 4014–4022.